\documentclass[journal]{IEEEtran}
\usepackage{graphicx}
\usepackage{amsfonts}
\usepackage{blindtext}
\usepackage{array}
\usepackage{cite} 
\usepackage{url}
\usepackage{xcolor}
\usepackage{soul}

\ifCLASSINFOpdf
\else
\fi
\begin{document}
%
\title{A methodology to evaluate the evolution of networks using topological data analysis}
\author{Joao Pita Costa * ** and Tihana Galinac Grbac * 
\thanks{
* University of Rijeka, Croatia, 

** Quintelligence, Slovenia 

\medskip

\textbf{Category:} G.2.2: Mathematics of Computing: Discrete Mathematics —
applications

}
}

\maketitle

\begin{abstract}
Networks are important representations in computer science to communicate structural aspects of a given system of interacting components. The evolution of a network has several topological properties that can provide us information on the network itself. In this paper, we present a methodology to compare the the topological characteristics of the evolution of a network, encoded into a (persistence) diagram that tracks the lifetimes of those features. This will enable us to classify the evolution of networks based on the distance between the diagrams that represent such network evolution. In that, we also consider complex vectors that bring a complementary perspective to the distance-based classification that is closer to the computational methods, aims to enhance the computational efficiency of those comparisons, and that is by itself a source of open research questions.
\end{abstract}

%
\IEEEpeerreviewmaketitle

\section{Introduction}

\subsection{Comparing the topology of the evolution of networks}

\begin{figure}
\centering
\includegraphics[height=3.8cm]{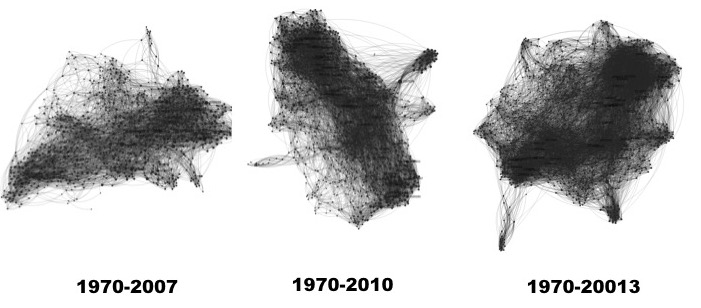}
\caption{The evolving network ScienceAtlas of the collaborations in scientific works by Slovenian researchers, evolving over a 9-year period with 3-year leaps. Each node represents an author and each edge represents a collaboration. The nodes with degrees smaller than 20 are filtered out so that the networks are not too large to be visualized.}
\label{evolvnet}
\vskip -6pt
\end{figure}

Networks that change as a function of time - known as evolving networks - are a natural extensions of undirected graphs (i.e., standard (static) networks). Almost all real world networks evolve over time, either by adding or removing nodes or edges. 
The example of scientific collaboration analysis, such as in the example of Figure \ref{evolvnet} shows such a network.

The analysis of the evolution of a network is a matter of interest transversal to many fields of knowledge, from social network analysis and scientific collaboration to computational biology. 
A standard example is the network dynamics of a social network such as Twitter should consider an evolution through time where new nodes come up as new members join, and new edges are created mirroring the new relationships between members that appear \cite{ba07}. Often all of these processes occur simultaneously in social networks.

Collaborative networks are a prime example of evolving networks, where nodes represent authors and edges represent scientific collaborations. This is illustrated in Figure \ref{evolvnet}. It shows the plot of three phases of an instance in the scientific community in Slovenia \cite{scat2} using ScienceAtlas, a web portal available at \url{scienceatlas.ijs.si} integrating data about 35272 researchers, 5716 projects, 82905 publications and 17190 video lectures. This too allows visualizing collaboration and competences of the researchers \cite{scat1}.

A biological network, on the other hand, is an approximate mathematical representation of connections found in ecological, evolutionary, and physiological research, among others. An example of a relevant application of such analysis of biological networks with respect to human diseases is network medicine. It considers networks in biological systems containing many components connected within complicated relationships but organized by simple principles \cite{ba07}.

\begin{figure}
\centering
\includegraphics[height=2.5cm]{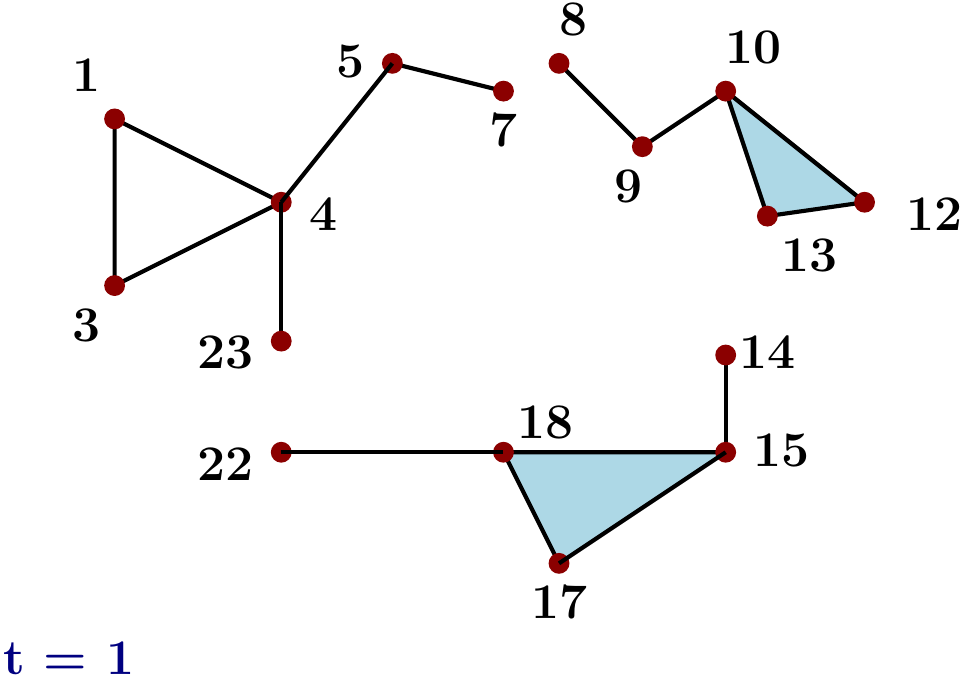}
\includegraphics[height=2.5cm]{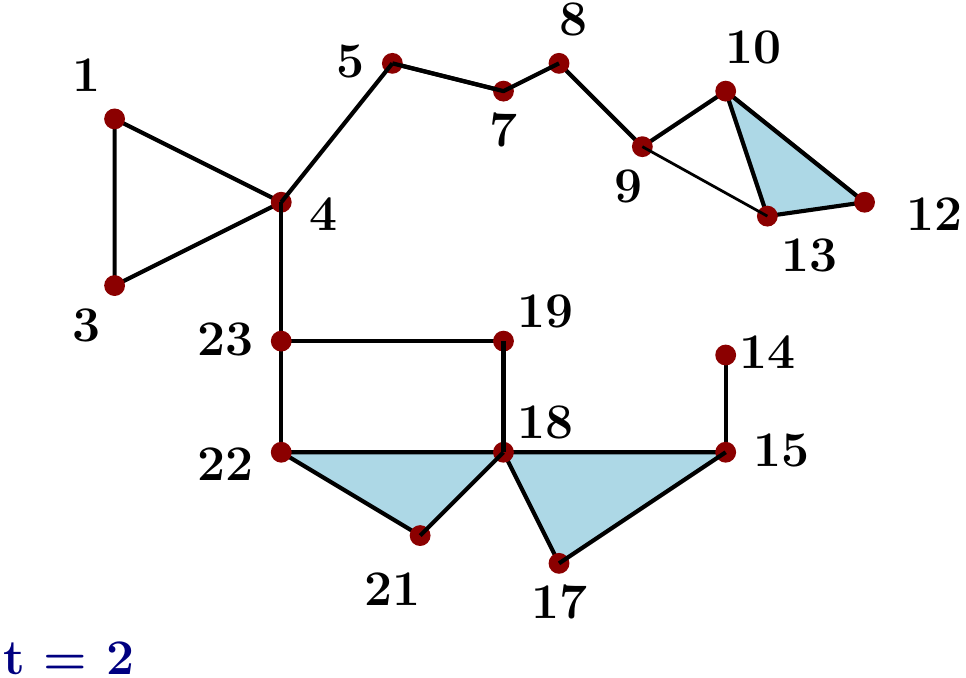}
\includegraphics[height=3cm]{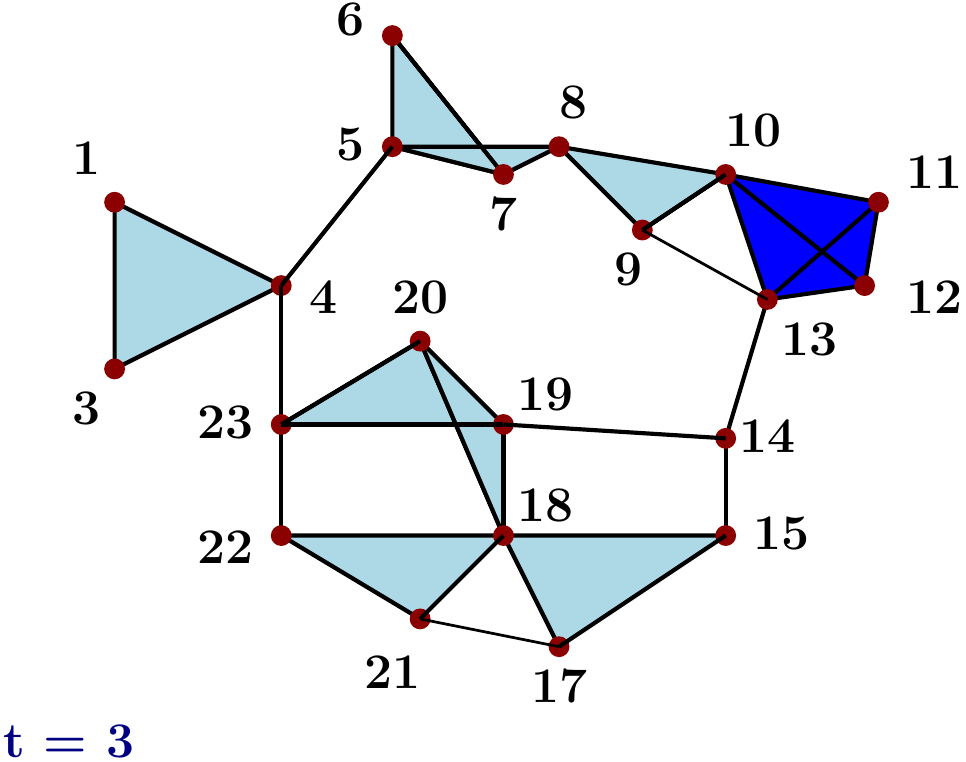}
\includegraphics[height=3cm]{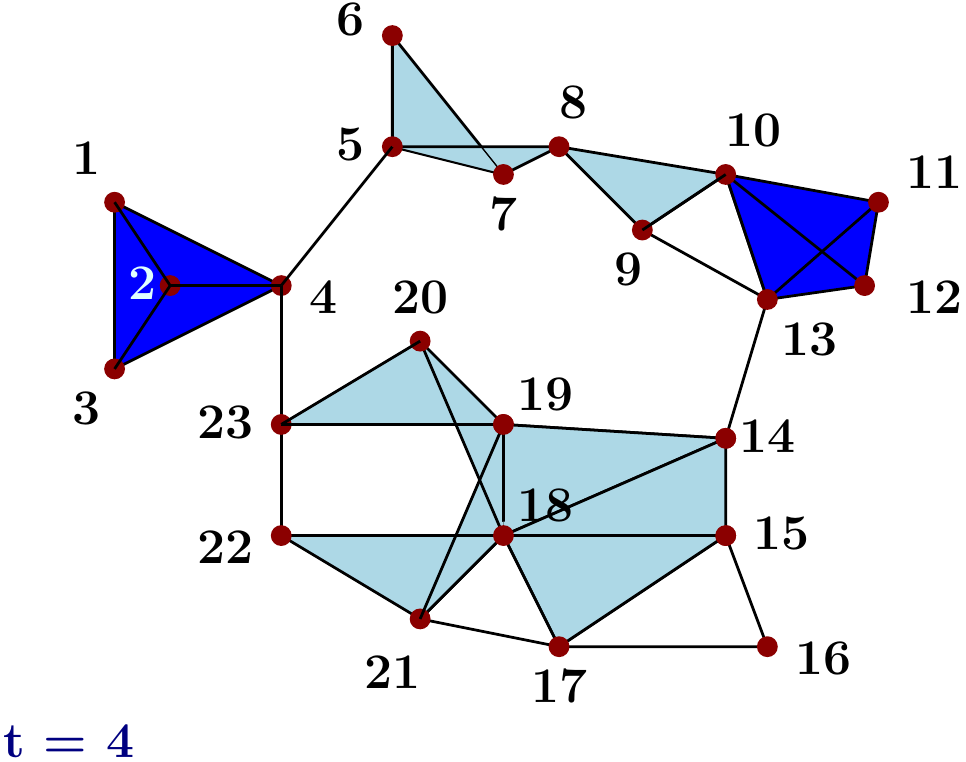}
\caption{An example of an evolving network based on an undirected graph that is growing over time by adding new edges and nodes at each step.}
\label{evolvnet3}
\vskip -6pt
\end{figure}

In this paper, we focus on the comparison of the evolution of two (or more) given networks. Our approach considers topological data analysis (TDA), allowing us to encode the topological features of the corresponding evolving networks onto diagrams, and using standard methods to compute distances between them. In that, we can  classify networks according to the distance between the topology of their evolution.

The TDA approach to the study of networks is not itself new. It had several widespread applications from collaboration networks \cite{CH13} to functional brain networks \cite{LCK11}. There are several ways of considering a height function in a network including: (i) considering weights in the edges of the network - \emph{weighted network} - and then having the function built by threshold those weights \cite{PSDV13}; (ii) measuring the distance from each node to each other by counting the minimal number of edges between them and then building the height function based on that distance \cite{HMR09}; among others. This permits us to use persistent homology over such height function. 
Another possibility is to consider the maximal cliques as the simplicial complexes (named \emph{clique complexes}) that feed the persistence algorithm and proceed with the computation directly over that \cite{EH10}. We used the latter approach to compute the persistence of the networks generated for the purpose of this paper.

\subsection{Basic notions in persistent homology}

\begin{figure}
\centering
\includegraphics[height=2.8cm]{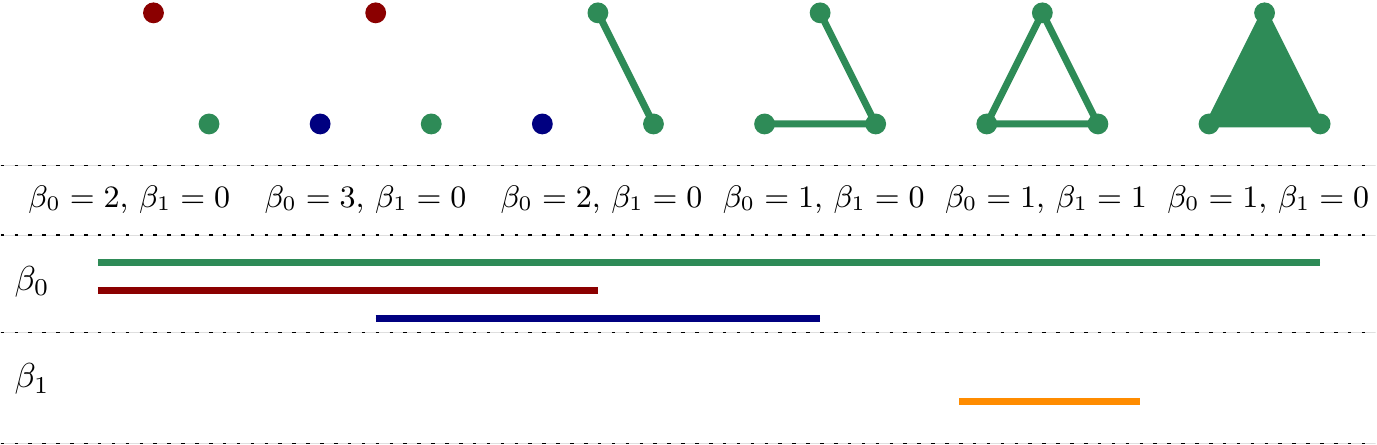}
\caption{The computation of persistent homology on a simplicial complex changing in time \cite{CH13}. The colors correspond to the topological features to which the lifetime is tracked in the persistence barcode below. The Betti numbers indicate the number of connected components $\beta_0$ in dimension zero, the number of holes in the network $\beta_1$ in dimension one, and the number of tunels and voids $\beta_2$ in dimension two.}
\label{evolvnet6}
\vskip -6pt
\end{figure}

Topology is a field of study in mathematics concerned in the quality aspects of an object. It focus on the properties that are preserved through deformations, twistings, and stretchings of the given continuous objects (e.g. linear maps) in multidimensional scenarios. Computational topology takes advantage of simplification methods (e.g. the triangulation of a space) to permit the computation of topological invariants.
One of those computations is homology which evaluates the connectedness of, e.g., a network at different dimensions separately. Thus, homology is a natural choice when it comes to the study of the topology of a network.
Now, if we consider a monotone function describing the time variable in, e.g., an evolving network, we can track its homology changes. This notion is known as persistent homology and is rooted in TDA, allowing for retrieving the essential topological features of an object \cite{car09}. 
Formally, persistent homology computes the topological features of a growing sequence of spaces $\emptyset = X_0\subseteq X_1 \dots \subseteq X_n = X$, known as a \emph{filtration} of the space $X$. $H_i(X)$ is the $i$-th homology group of $X$, with an associated $i$-th Betti number of $X$,$\beta_i$, corresponding to the measure of connectedness in the $i$-th dimension (cf. \cite{EH10}).
Using the inclusion maps $X_j \to X_{j+1}$ we can identify copies of $Z_2$ in the homology groups $H_i(X_j)$ and $H_i(X_{j+1})$ of a filtration and track where the homology changes. We do that by  recording when a new copy appears (i.e. "is born"), and when an existing copy persists or merges to an existing one (i.e. "dies"). That persistence of the topological feature is tracked by a lifetime bar (as shown in Figure \ref{evolvnet6}) that can be equivalently represented by an ordered pair $(x,y)$, where $x$ is the birth time and $y$ is the death time. The multiset of all such points exists in the plane subset defined by $0<x<y$ that encodes the topology of a space and is known as \emph{persistence diagram}. Several topological features can have the same lifetimes and therefore some of the points in the persistence diagram are repeated in the multiset. We refer to their amount as \emph{multiplicity}. We consider the infinite points in the diagonal as points of the persistence diagram with null lifetime.  
The standard method to compare two persistence diagrams - called \emph{bottleneck distance} - measures the cost of finding a correspondence between their points. It identifies the closest matching elements of each persistence diagram and determines the global distance based on what is the biggest of those distances. The cost of taking a point $p = (p_1, p_2)$ to a point $q =(q_1, q_2)$ in $R^2$ is given by the $L_{\infty}$ norm $\| p - q\|_\infty = max{|p_1-q_1|,|p_2-q_2|}$. Then, the bottleneck distance between persistence diagrams $X$ and $Y$ is computed by taking the infimum over all such matchings, i.e., $d_B(X, Y) = \inf_\eta \sup_{x\in X} \|x - \eta(x)\|_\infty$,
where the infimum is taken over all bijections $\eta$ from $X$ to $Y$. Each point with multiplicity $k$ in a multiset is interpreted as $k$ individual points, and the bijection is interpreted between the resulting sets \cite{ed08}. 

\subsection{The motivation of EVOSOFT}
Nowadays, software systems start to interconnect to provide new and innovative applications and services that drives new development opportunities in all domains. Therefore these software systems have gradually evolved into large scale complex systems and we lack models for their further management and evolution. One of the key aspects of such systems is the ability to model and predict their behaviour to achieve the required quality of operations to fulfill human expectations in all domains. 
In that, the project Evolving Software Systems: Analysis and Innovative Approaches for Smart Management (EVOSOFT) aims to understand how abstract software structures can be used to model global system properties (e.g. fault distributions). Understanding how to use software structure to model fault distributions can help us to improve system reliability. EVOSOFT observes software structure as networks with nodes representing various software functions that are interconnected to each other by function calls. In particular, a software graph structure considers nodes as program functions (e.g. classes in object oriented paradigm, functions or modules in functional programming) and edges as function calls or signals transferred in communication among these program functions. EVOSOFT aims to observe how large software systems evolve from version to version, and understand the relationship between the change in software structure during its evolution, and the change in software fault distributions across its structure. Previous empirical studies in \cite{GalinacGrbacTSE, GalinacGrbacIST, Petric} show that communication structures among the program functions significantly influence system fault distributions. This is what motivated us to further explore this relationship.

\section{Use-case methodology to encode and compare evolving networks}

\begin{figure}
\centering
\includegraphics[height=1.3cm]{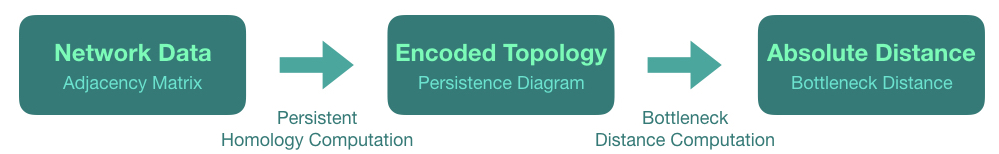}
\caption{The methodology diagram to encode and compare the topology of the evolution of two (or more) networks using TDA. It considers three phases: (i) the data, where we input the evolving network represented by an adjacency matrix; (ii) the topology, where a persistence diagram encodes the topological invariants of the network evolution; and (iii) the distance, held between persistence diagrams representing the topology of given evolving networks.}
\label{pipeline}
\vskip -6pt
\end{figure}

The problem of tracking and comparing the evolution of networks can be very demanding and complex due to the combinatorial properties of networks.
In the following section we shall describe the methodology diagram to encode a compare the topology of the evolution of networks (as illustrated in Figure \ref{pipeline}). 
It considers persistent homology to encode the topological features of the evolution of a network using persistence diagrams. In that, we first provide the evolving network given by one Boolean adjacency matrix for each phase of network development. 
We then compute the persistent homology of the evolving network by feeding the concatenated matrices a suitable algorithm. It will encode the topology of each evolving network, representing it by one unique persistence diagram each. 
Finally, we measure the bottleneck distance between persistence diagrams to identify how close are the evolving networks to each other based on their topology. 

\begin{figure}
\centering
\includegraphics[height=2cm]{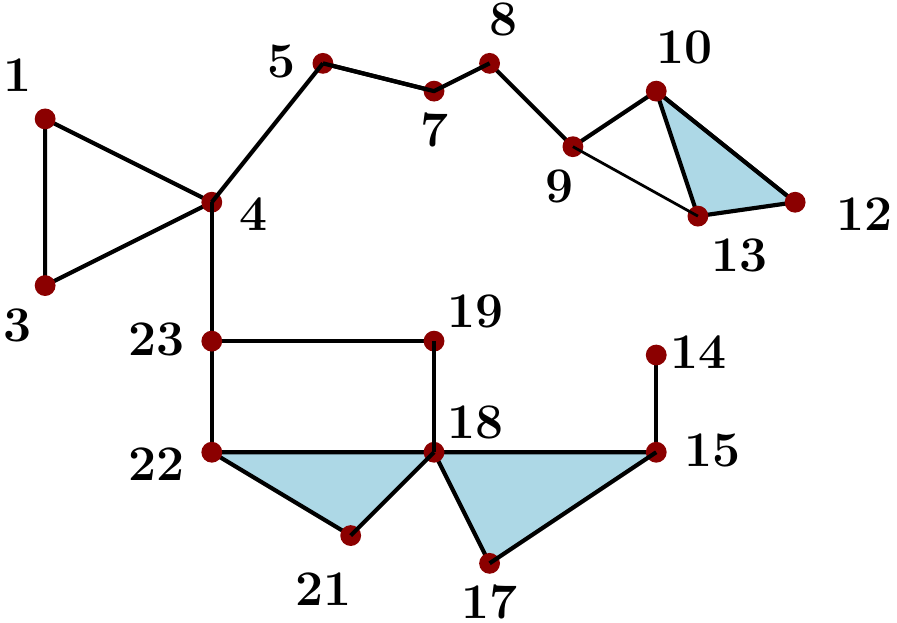}
\includegraphics[height=2cm]{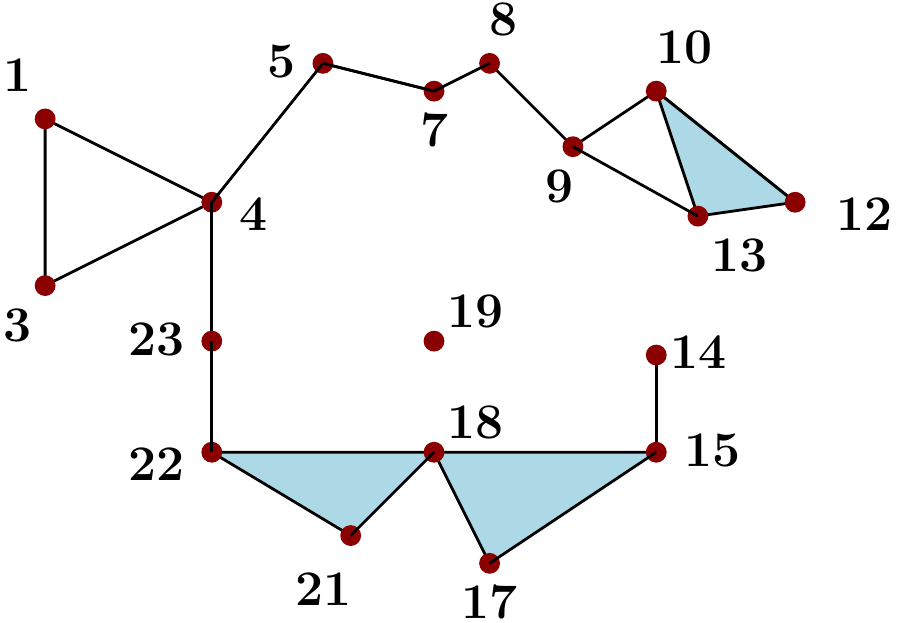}
\includegraphics[height=2cm]{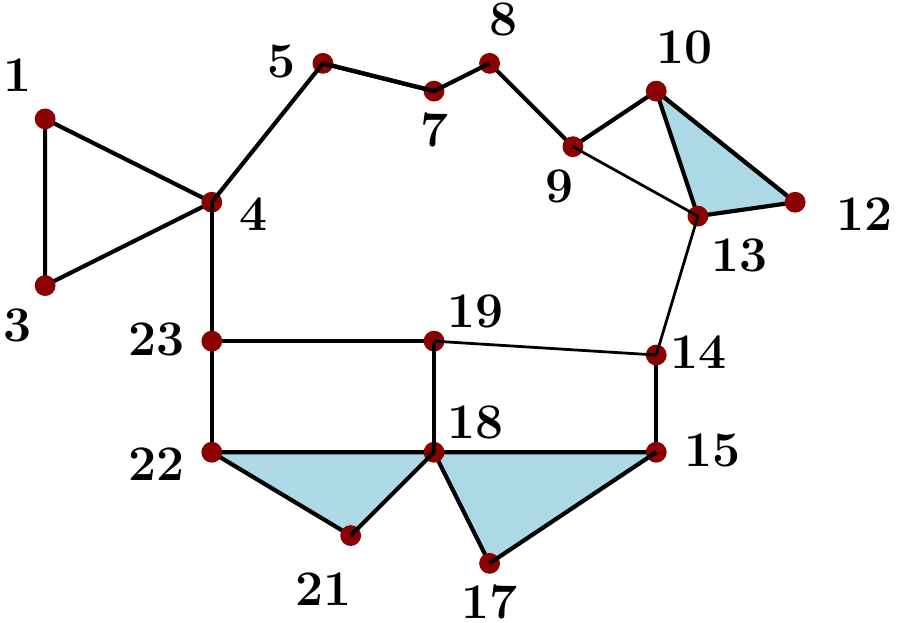}\\
\vspace{0.5cm}
\includegraphics[height=2cm]{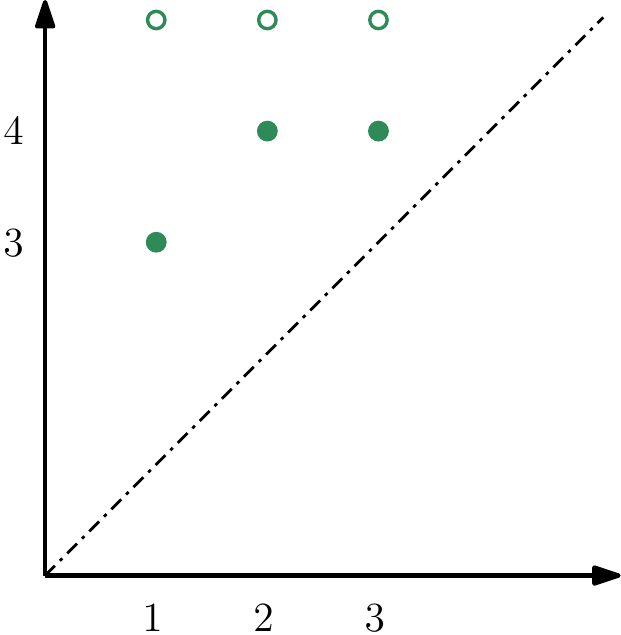} \hspace{0.8cm}
\includegraphics[height=2cm]{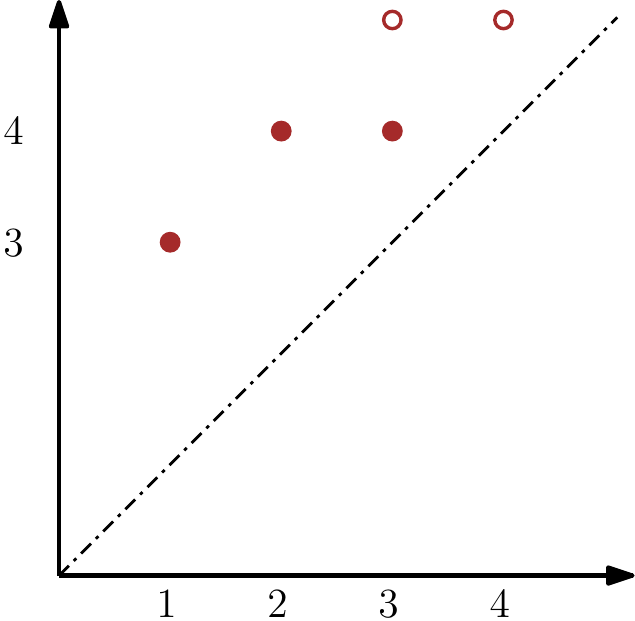} \hspace{0.8cm}
\includegraphics[height=2cm]{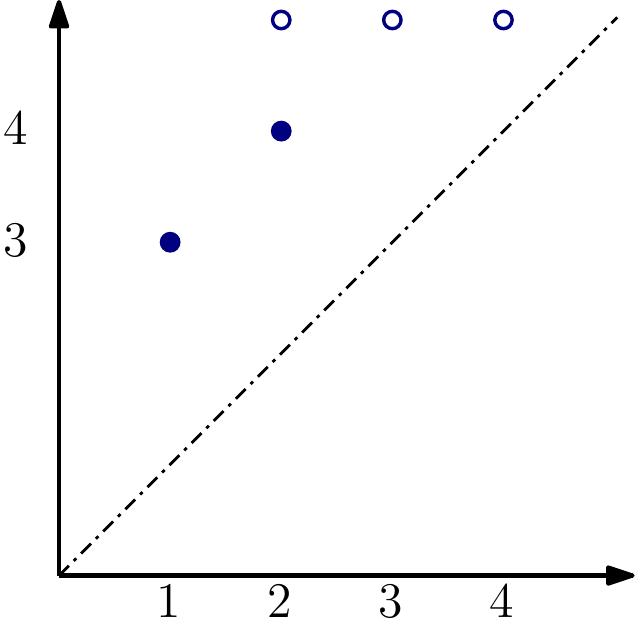}\\
\caption{The presented methodology applied to the comparison of the evolution of three networks sharing the same evolution as the network represented in Figure \ref{evolvnet3} but with differences in phase 2. Each evolving network is associated with one persistence diagram that encodes its topology. This permits us to visualize the relevant topological features of the evolution of the networks. 
}
\label{evolvnet2}
\vskip -6pt
\end{figure}

To the purpose of this paper, we used the software library \emph{Perseus} \cite{perseus} to compute the homology of a the evolving network represented in Figure \ref{evolvnet3}, given by the graph's Boolean adjacency matrix. The network is provided to \emph{Perseus} as a list of cliques including the time of appearance. The output of that procedure is a persistence diagram that corresponds to the topological changes within the evolution of that network. The evolving network $A$ on the left has four stages as illustrated in Figure \ref{evolvnet3}. The evolving networks $B$ and $C$ are variations of the evolution of the end network in $A$ with different phases at time $t=2$, as represented in \ref{evolvnet2}. 

To compare the evolution of networks we consider the distance between the corresponding persistence diagrams, using the bottleneck distance. This permits a fast computation of the distance between the (persistence diagrams representing the) topology of two evolving networks. In the case of the persistence diagrams encoding the topology of evolving networks A, B and C represented in Figure \ref{evolvnet2}, we get $d(A,B)=0$ and $d(B,C)=d(A,C)=1$. 
This discards the points with infinite persistence that are less relevant when considering dimension 1 diagrams.
The computations were done using the TDA package available in R \cite{RTDA2}.
In this example we can explore the distance between several possible evolution of a network. In it, shows how TDA can contribute to better understand the behavior of a certain network.

\section{The EVOSOFT Experiments}

For the purpose of this research we will use the EVOSOFT motivation to generate networks that fit that scenario and allow us to compare the evolution of networks in that context. In these preliminary experiments we shall consider data representing the evolution of networks based on the empirical analysis of the evolution of complex software systems \cite{GalinacContel}. 


In these experiment we will generate networks with labeled nodes - not ordered pairs in $\mathbb{R}^2$ - and extract all maximal cliques from it. The maximal cliques serve us to construct clique complexes with which we are able to later on compute the topology of those networks. 
In these experiments we shall obtain the EVOSOFT evolving networks provided by their graph's Boolean adjacency matrix. Those matrices must be consistent with the evolution of the network in the sense that existing maximal cliques in phase $i$ must maintain or enlarge in the phase $i+1$ during the updates of a software version.  
The persistence diagrams computed by \emph{Perseus} shall exhibit the encoded topology of evolving networks corresponding to different pieces of software.


The comparison between the topology of a pair of evolving networks given by the adjacency matrix is given by the bottleneck distance between the corresponding diagrams. That distance can be computed using the R library \cite{RTDA2}. When considering other evolving networks we can calculate the pairwise distance between all of them and consider single linkage clustering based on this metric (as in earlier TDA applications to gene expression data as in \cite{MJ2014}) to allow classification based on the topology of network evolution.

\section{Comparison through complex vectors}

A possible algebraic representation of persistence diagrams is offered by complex polynomials. The method layed out in \cite{Bar15} can lead to avoid tedious and less meaningful computations of bottleneck distance, since far polynomials represent far persistence diagrams (the converse is known not to be true). A fast comparison of the coefficient vectors can reduce the size of the database to be classified by the bottleneck distance. We can then focus on close persistence diagrams for which we want to calculate precise measures. This should complement existing methods, rising the efficiency of computations for large evolving networks. 
Given a persistence diagram $D$ described by its points $p_1 = (u_1, v_1), \dots , p_s = (u_s, v_s)$ with multiplicities $r_1,\dots,r_s$, respectively, the method considers complex numbers $z_1=u_1+iv_1,\dots, z_s=u_s+iv_s$. This allows us to associate to $D$ the complex polynomial $fD(t) = \Pi^s_{j=1}(t − z_j)^{r_j}$ where $r_j$ is the multiplicity of the point $p_j$. 
It was shown in \cite{Bar15} that the first $k$ coefficients are the ones carrying most of the relevant information and, therefore, the choice of a threshold $k$ can reduce the computational complexity. 

The unpublished 2-part algorithm by the authors of \cite{Bar15} permits us to input a persistence diagram in order to compute a complex vector out of it. Then the same algorithm compares two complex vectors corresponding to two persistence diagrams to output a float corresponding to the distance between those vectors. At the moment, this approach to convert persistence diagrams into complex vectors can be applied only when neglecting points with infinite persistence.
In the running example we get the polynomial $p_A=(t-1-3i)(t-2-4i)(t-3-4i)=p_B$ and $p_C=(t-1-3i)(t-2-4i)^2$, not considering points of infinite persistence.
We then develop the polynomials to identify their coefficients into a complex vector. The distance between the three complex vectors corresponds to a basic classification of the given evolving networks.  
This is not a dense case where we would need additional tools like complex vectors. Though, real life examples of evolving networks are appropriate cases of such needs due to their inherent complexity.

\section{Conclusions and further work}

In this paper we have discussed the topological data analysis of evolving networks. In that we presented a method to encode the topology of the evolution of a given network through a persistence diagram, and its potential for a classification based on a chosen distance between diagrams. 
The inherent complexity of an evolving network demands for the data simplification methods to be available and appropriate to the nature of the considered object. In that, the TDA-based methodology in this paper can contribute to the analysis and interpretation of evolving networks and their behaviour. The experiments in real data are valuable to improve this method. In that, the collaborations with the earlier mentioned \emph{Slovenian Science Atlas} would be welcome, allowing us to further explore the interpretation of the topology of the evolution of these collaborative networks and the distance between them. 
Further work includes the processing of EVOSOFT existing networks, as well as the interpretation of results in the context of that field of knowledge. It can provide new challenges specific to the available data and to its role and usage in the field. In particular, the interpretation of the persistent topological features captured in EVOSOFT experiments represents a relevant open problem that requires a deeper analysis based on the EVOSOFT expertise and the manipulation of the topological results.
Lastly, the mathematical development of the complex vector method, that contributes to the study of evolving networks in general, is a rather computational method that is suitable to the application of compatible algorithms, allowing potential engineering applications. Moreover, it is itself a great source of open mathematical problems that we shall consider in further research (e.g. stability \cite{EH10}).

\section*{Acknowledgment}
The authors would like to thank to Barbara di Fabio for the usefull discussions on complex vectors and advice on further research, and to Primo\v{z} \v{S}kraba for his comments and suggestions. 
The first and second authors would like to thank to the support of the Croatian Science Foundation's funding of the project EVOSOFT (UIP-2014-09-7945). The second author would also like to thank to the support by the University of Rijeka Research Grant 13.09.2.2.16 funding.

\ifCLASSOPTIONcaptionsoff
  \newpage
\fi

\end{document}